\documentclass[preprint,aps,pra,showpacs,floats,superscriptaddress]{revtex4}
\usepackage[dvips]{graphicx}
\usepackage{amsmath,bm}
\usepackage{psfrag}

\begin{document}

\title{Dewetting of an ultrathin solid film on a lattice-matched or amorphous substrate}

\author{M.~Khenner}
\affiliation{Department of Mathematics, State University of New
York at Buffalo, Buffalo, NY 14260, USA}

\newcommand{\Section}[1]{\setcounter{equation}{0} \section{#1}}
\newcommand{\rf}[1]{(\ref{#1})}
\newcommand{\beq}[1]{ \begin{equation}\label{#1} }
\newcommand{\eeq}{\end{equation} }

\begin{abstract}

An evolution partial differential equation for the surface of a non-wetting single-crystal film
in an attractive substrate potential is derived and used to study
the dynamics of a pinhole for the varying initial depth of a pinhole and the strengths of the potential and 
the surface energy anisotropy. 
The results of the simulations demonstrate how 
the corresponding parameters
may lead to complete or partial dewetting of the film.
Anisotropy of the surface energy, through faceting of the pinhole walls, is found 
to most drastically affect the time to film rupture. In particular, the similations 
support the conjecture that the strong anisotropy is capable of the complete suppression of dewetting
even when the attractive substrate potential is strong.
\end{abstract}

\date{\today}
\maketitle

\begin{center}
{\bf I. INTRODUCTION}
\end{center}

Under typical operation conditions
the thin solid film devices heat up to temperatures 
as high as 1000$^\circ$, without melting.
A high temperature activates the mass transport by surface diffusion. Thus the film morphology changes, which often results
in
harmful impacts on a device structural integrity. On the other hand, it is well known that thin \emph{liquid} films 
(where the mass transport is by bulk hydrodynamic flows) are subject to 
dewetting caused by the long-range film-substrate interactions (which are also called wetting interactions). Dewetting results in 
film rupture and the agglomeration of liquid in droplets. 
The wetting interactions contribute noticably to the liquid film dynamics when the thickness of a film is of the 
order of few microns. Since this characteristic length scale is of the order of ten nanometers for solid films
\cite{YZSLLHL,SECS},
and due to a high temperature activation barrier for surface diffusion, the dewetting of solid thin films is not as
common. The structural integrity of a film must be maintained for a device to function properly.
Thus it is important to understand the process of dewetting and find means to control it.

One material system which exhibits solid film dewetting is the silicon-on-insulator. 
It comprises the ultrathin silicon film deposited on the amorphous SiO$_2$ substrate.
The main feature of such system is the absence of stress at the film-substrate interface.
The experiments \cite{YZSLLHL,SECS} with the sub-10nm Si films at 800-900 $^\circ$C demonstrate that 
dewetting starts at pinholes
in the Si planar surface. 
The pinholes may exist prior to the annealing, or they form shortly after the temperature is raised.
The critical depth of the
pinhole which is necessary to initiate dewetting is presently unknown (apparently, not all pinholes dewet), 
nor is known the kinetics of the 
pinhole deepening and the pinhole shape. 

Due to anisotropy of the film surface energy
it is natural to expect that the dynamics of a pinhole and the rupture process are qualitatively
different from dewetting and rupture of the liquid surface having an isotropic surface tension. In particular,
since the crystal surface tends to facet when its orientation falls in the unstable range, 
it is not \emph{a priori} clear that dewetting, when it is
initiated by substrate attraction of the initially slightly deformed free surface, 
not exhibiting unstable orientations, will proceed to rupture. Somewhat similar situation occurs
in strongly anisotropic grain boundary grooving. Interestingly, it was demonstrated \cite{XinWong03} 
that such groove can be smooth and have the same shape as the isotropic groove, but the growth
rate is reduced by a factor that depends on the degree of anisotropy.

This paper attempts to clarify the isssue by studying the 
dynamics of the pinhole with the varying initial depth, and for different strengths of the potential and 
the surface energy anisotropy. 

There have been several theoretical models published on the dynamics of thin solid films, where 
wetting interactions are accounted for
\cite{ORS}-\cite{Chiu};
these continuum models are primarily designed for the description of a self-assembly process
in the course of the epitaxial 
crystal growth. In contrast to this paper, these models assume that the film wets the substrate,
meaning that $\partial \mu_w/\partial h|_{h=const.} > 0$, where $\mu_w$ is the wetting chemical potential,
and $h$ is the film height.
Also, dewetting of a solid monolayer was recently studied \cite{PL} using a solid-on-solid model
and the Kinetic Monte-Carlo simulations. Dewetting of a monolayer proceeds through the nucleation 
and growth of holes in the film, which are caused by the discreteness of the underlying crystalline
lattice and the substrate roughness. 
This is not the case for thicker films considered here. 

For solid films there is currently several choices of the wetting model, including the two-layer wetting 
model \cite{ORS,CG,GLSD,LGDV} and the van der Waals potential \cite{SZ,Chiu}. In this study the former model is used
due to its generality.
It has been demonstrated in Refs. \cite{T,BWA} that
the two-layer wetting model is consistent with \textit{ab initio} calculations, but
presently there is no experimental studies in which the wetting interaction potential has been measured.

The continuum model of this paper is based on the well-established
physico-mathematical framework which originates from the works of Mullins \cite{MULLINS5759,MULLINS_Metals} and Herring 
\cite{HERRING,HERRING1}
half a century ago. Mullins derived the first partial differential equations for the description
of the surface morphology changes by surface diffusion (capillarity), evaporation-recondensation and bulk
diffusion. Herring was first to systematically study the dependence of the
equilibrium crystal shape \cite{Wulff} on the surface energy anisotropy. 
The experiments \cite{Bonzel_physrep} demonstrate that the equilibrium shape
can be partially or completely faceted, with the atomically sharp edges and corners and straight facets
emerging
 when the anisotropy
is strong, or with the smooth junctions and curved facets emerging when the anisotropy is weak.
Other authors  studied the dynamics of the
crystal surface evolution with the strongly anisotropic surface energy 
and ways of regularizing the mathematical 
ill-posedness  of the evolution problem (and the removal of the associated fine-scale instability) 
\cite{XinWong03}, \cite{LiuMetiu} - \cite{WN}. The preferred method of regularization is
the addition of the small curvature-dependent term to the surface energy 
\cite{AG} - \cite{Brian_regular}, \cite{LiuMetiu,GoDaNe98}. 
This term penalizes the sharp corners and makes them rounded on a small length scale \cite{HERRING1}.

\begin{center}
{\bf II. PROBLEM STATEMENT}
\end{center}

A two-dimensional film with the free one-dimensional surface $z=h(x,t)$ is assumed, where
$h$ is the height of the film surface above the substrate. The surface
evolves by surface diffusion.

Under these assumptions, the governing equation for $h(x,t)$ has the form
\begin{equation}
h_t = \frac{\Omega D \nu}{kT}\frac{\partial}{\partial x}\left((1+h_x^2)^{-1/2}\frac{\partial \mu}{\partial x}\right),
\label{1.1}
\end{equation}
where $\Omega$ is the atomic volume, 
$D$ the adatoms diffusivity, $\nu$ the adatoms surface density, $k$ the Boltzmann constant,
$T$ the absolute temperature, and $\mu = \mu_{\kappa}+\mu_w$ the surface chemical potential.
Here $\mu_{\kappa}$ is the regular contribution due to the surface mean curvature $\kappa$ \cite{MULLINS5759,TCH94}.
Also $(1+h_x^2)^{-1/2}=\cos{\theta}$,
where $\theta$ is the angle that the unit surface normal makes with the [01] crystalline direction, along which is
the $z$-axis.
(The $x$-axis is along the [10] direction.)
Thus $\theta$ measures the orientation of the surface with respect to the underlying crystal structure.
Note that here and below the subscripts $x, t$ and $\theta$ denote differentiation.

The wetting chemical potential 
\begin{equation}
\mu_w = \Omega \left(1+h_x^2\right)^{-1/2}\frac{\partial \gamma(h)}{\partial h},
\label{1.4e}
\end{equation}
where, in the two-layer wetting model, 
\begin{equation}
\gamma(h) = \gamma^{(f)}(\theta) + \left(\gamma_s-\gamma^{(f)}(\theta)\right)\exp{\left(-h/\ell\right)}
\label{1.4f}
\end{equation}
is the thickness-dependent surface energy density of the film-substrate interface. 
Here $\gamma_s=const.$ is the surface energy density of the substrate in the absence
of the film, and $\ell$ is the characteristic wetting length. Thus $\gamma(h) \rightarrow \gamma^{(f)}$ 
as $h\rightarrow \infty$, and $\gamma(h) \rightarrow \gamma_s$ as $h\rightarrow 0$.
$\gamma^{(f)}(\theta)$ is the energy density of the film surface, assumed
strongly anisotropic. The strong anisotropy 
possibly leads to faceting. (Below, when a reference to the surface energy density is made, the word ``density" is omitted for brevity).

$\gamma^{(f)}(\theta)$ is taken in the form 
\begin{equation}
\gamma^{(f)}(\theta) = \gamma_0 (1+\epsilon_\gamma \cos{4\theta}) + \frac{\delta}{2}\kappa^2,
\label{1.2}
\end{equation}
where $\gamma_0$ is the mean value of the film surface energy
in the absence of the substrate
potential (equivalently, the surface energy of a very thick film), $\epsilon_\gamma$ determines the degree of anisotropy,
and $\delta$ is the small non-negative regularization parameter having units of energy. 
The $\delta$-term in Eq. \rf{1.2} makes the evolution equation \rf{1.1} mathematically well-posed for strong 
anisotropy \cite{AG} - \cite{Brian_regular}, \cite{LiuMetiu,GoDaNe98}. 
(The anisotropy is weak when $0<\epsilon_\gamma<1/15$ and strong when $\epsilon_\gamma\ge 1/15$.
$\delta=0$ in the former case. In the latter case the polar plot of $\gamma^{(f)}(\theta)$ has cusps at the orientations
that are missing from the equilibrium Wulff shape and the surface stiffness 
$\gamma^{(f)}+\gamma^{(f)}_{\theta\theta}$ is negative
at these orientations \cite{HERRING,HERRING1}. Thus the evolution equation is ill-posed unless regularized 
\cite{AG,CGP}.)
The form \rf{1.2} assumes that the surface energy is maximum in the [01] direction.

The curvature contribution to the chemical potential has the form 
\begin{equation}
\mu_{\kappa} = 
\Omega (\gamma^{(f)}+\gamma^{(f)}_{\theta \theta} - H)\kappa, 
\label{1.4b}
\end{equation}
where $\gamma^{(f)}(\theta)$ has the form \rf{1.2} with the mean value $\gamma_0$ replaced by the effective
value $\bar \gamma_0$ which reflects the presence of the substrate:
\begin{equation}
\bar \gamma_0 = \frac{1}{h_0}\int_0^{h_0} \left[\gamma_0 + \left(\gamma_s-\gamma_0\right)\exp{\left(-z/\ell\right)}\right] dz.
\label{mean}
\end{equation}
Here $h_0=const.$ is the unperturbed film height. The $H$-term in Eq. \rf{1.4b} is non-zero only for strong
anisotropy, 
and it ensures the continuity of the chemical potential across corners on the faceted crystal shape.
The form of this term is \cite{GoDaNe98}
\begin{equation}
H = \kappa\left(\frac{1}{2}+\frac{d^2}{d\theta^2}\right)
\left(\frac{\partial \gamma^{(f)}}{\partial \kappa}\right)+\frac{\partial^2 \gamma^{(f)}}{\partial \kappa^2}
\left(\frac{d\kappa}{d\theta}\right)^2.
\label{1.4bb}
\end{equation}
%
%
Finally, $\kappa = -h_{xx}(1+h_x^2)^{-3/2}$.

To non-dimensionalize the problem, $h_0$ is chosen as the length scale,
and $h_0^2/D$ as the time scale. Also, let $r = \ell/h_0$. 
Substitution of $\mu_\kappa$ and $\mu_w$ in Eq. \rf{1.1} gives
\begin{equation}
h_t = B\frac{\partial}{\partial x}\left(q P^{(1)}_\kappa-\Delta P^{(2)}_\kappa+P^{(1)}_w-\Delta P^{(2)}_w\right),
\label{1.5}
\end{equation}
where $h,\ x$ and $t$ are non-dimensional, $B = \Omega^2\nu \gamma_0/(kTh_0^2)$, 
$\Delta = \delta/(\gamma_0h_0^2)$ is the non-dimensional regularization parameter,
and $q$, $P^{(1,2)}_\kappa$ and $P^{(1,2)}_w$ are the following quantities.
\begin{equation}
q = 1+r\left(\frac{\gamma_s}{\gamma_0}-1\right)\left(1-\exp{(-1/r)}\right) > 0,
\label{1.6a0}
\end{equation}
\begin{equation}
P^{(1)}_\kappa = \frac{\partial}{\partial x}\left[\left(1-15\epsilon_\gamma \cos{4\theta}\right)\kappa\right] \cos{\theta}; 
\label{1.6a}
\end{equation}
\begin{equation}
P^{(2)}_\kappa = \frac{\partial}{\partial x}\left[\left(\frac{1}{2}\kappa^3+
\cos{\theta}\frac{\partial}{\partial x}\left(\cos{\theta}\frac{\partial \kappa}{\partial x}\right)\right)\right] \cos{\theta};
\label{1.6b}
\end{equation}
\begin{eqnarray}
P^{(1)}_w &=& \frac{\exp{\left(-h/r\right)}h_x}{r(1+h_x^2)^4}\left[r^{-1}a_1\left(1+h_x^6\right)+
\right. \nonumber \\ &&  \left. 3r^{-1}a_5h_x^2\left(1+h_x^2\right)+a_2+2a_4h_x^2+a_1h_x^4\right];
\label{1.6c}
\end{eqnarray}
\begin{eqnarray}
P^{(2)}_w &=& \frac{\exp{\left(-h/r\right)}}{2r(1+h_x^2)^5}\left[r^{-1}h_{xx}^2h_x\left(1+h_x^2\right)+\right. \nonumber \\
&& \left. h_{xx}\left(7h_{xx}^2h_x-2h_{xxx}\left(1+h_x^2\right)\right)\right].
\label{1.6d}
\end{eqnarray}
%
Note that $P^{(1)}_w$ and $P^{(2)}_w$ are proportional to $\exp{\left(-h/r\right)}$.
In Eqs. \rf{1.6c} and \rf{1.6d},
$a_1 = \Gamma - 1 - \epsilon_\gamma,\ a_2 = \Gamma - 1 - 17\epsilon_\gamma,\
a_4 = \Gamma - 1 + 11\epsilon_\gamma$ and $a_5 = \Gamma - 1 + 5\epsilon_\gamma/3$.
Thus Eq. \rf{1.5} contains five parameters, $B, r, \epsilon_\gamma, \Delta$ and $\Gamma = \gamma_s/\gamma_0$
(where $\Delta=0$ when $0<\epsilon_\gamma<1/15$).

From Eqs. \rf{1.4e}-\rf{1.2} it is easy to see that the requirement of zero wetting between the film 
and the substrate,
$\partial \mu_w/\partial h|_{h=h_0} < 0$, is equivalent to requiring negative $a_1$, i.e. $\Gamma < 1 + \epsilon_\gamma$. Thus in what follows $a_1<0$.

Eq. \rf{1.5} can be written also in the widely used small-slope approximation (SSA), where $|\partial/\partial x| = \epsilon \ll 1$: 
\begin{equation}
h_t = B\frac{\partial}{\partial x}\left(q P^{(1)}_\kappa-\Delta P^{(2)}_\kappa+P^{(1)}_w-\Delta P^{(2)}_w\right),
\label{1.7}
\end{equation}
\begin{equation}
P^{(1)}_\kappa = \Lambda_1h_{xxx}+\Lambda_2h_{xx}^2h_x+\Lambda_3h_{xxx}h_x^2; 
\label{1.8a}
\end{equation}
\begin{equation}
P^{(2)}_\kappa = h_{xxxxx};
\label{1.8b}
\end{equation}
\begin{eqnarray}
P^{(1)}_w &=& \frac{-\exp{\left(-h/r\right)}}{r}\left[a_2h_{xx}h_x\left(1-5h_x^2\right)+
\right. \nonumber \\
&& \left. r^{-1}h_x\left(a_1+\left(2a_3-3a_1\right)h_x^2+a_6h_x^4\right)\right];
\label{1.8c}
\end{eqnarray}
\begin{equation}
P^{(2)}_w = \frac{\exp{\left(-h/r\right)}}{r}h_{xx}\left[\frac{1}{2}r^{-1}h_{xx}h_x-h_{xxx}\right].
\label{1.8d}
\end{equation}
%
In Eq. \rf{1.8a} $\Lambda_1 = 15\epsilon_\gamma - 1,\ \Lambda_2 = 3-285 \epsilon_\gamma$ and
$\Lambda_3 = 2-150 \epsilon_\gamma.$ 
In Eq. \rf{1.8c} $a_3 = \Gamma - 1 + 3\epsilon_\gamma$ and $a_6 = \Gamma - 1 - 25\epsilon_\gamma$. Also note that the non-negative $\Lambda_1$ signals that the surface energy anisotropy is strong.
Eq. \rf{1.7} has the same form as the evolution equation (26) of Ref. \cite{GLSD}.

In the simulations reported below, the following values of the physical parameters are used:
$D=1.5\times 10^{-6}$ cm$^2$/s, $\Omega = 2\times10^{-23}$ cm$^3$, $\gamma_0 = 10^3$ erg/cm$^2$, 
$\gamma_s = 5\times10^2$ erg/cm$^2$, $\nu = 10^{15}$ cm$^{-2}$, $kT = 1.12\times10^{-13}$ erg, $h_0= 10^{-6}$ cm,
and $\delta = 5\times 10^{-12}$ erg. These values translate into $B=3.57\times10^{-3}$, $\Gamma = 0.5$ and 
$\Delta = 5\times 10^{-3}$ (or zero). Values of the parameters $\epsilon_\gamma$ and $r$  will be chosen later.

The Method of Lines is adopted for the computation.
Eqs. \rf{1.5} and \rf{1.7} are discretized
by finite differences on a spatially-uniform staggered grid in $x$. In Eq. \rf{1.5} the terms $P^{(1,2)}_\kappa$ 
are discretized
in the nested fashion \cite{MYPRE2}. This, in particular, requires the explicit computation of only the first- and 
second-order partial derivatives $h_x$ and $h_{xx}$. Thus, the numerical stiffness associated with the
higher order derivatives becomes less pronounced, which manifests in the significant
speed-up of the computation (in the cases of the large surface slope and faceting, 
the speed-up factor can be as large as 300). 
The results provided by this and by the standard scheme agree in all cases
where such a comparison was made. The integration in time of the resulting coupled system of the 
ordinary differential equations
is done using the implicit Runge-Kutta method.

The initial condition in all runs is the surface 
\begin{equation}
h(x,0) = 1+\alpha\cos{k_{max}x},
\label{ini_cond}
\end{equation}
where $\alpha$ is
the non-dimensional perturbation amplitude (assumed not necessarily small, since the initial pinhole may 
extend deep into the film), 
and $k_{max}$ is the wavenumber for which the linear growth rate
of the instability is the maximum. This wavenumber (or the corresponding wavelength $\lambda_{max}$) 
is usually referred to as ``the most dangerous mode". 
In Eq. \rf{ini_cond}, $0\le x\le \lambda_{max}$ (i.e., one wavelength).
The boundary conditions at $x=0,\ \lambda_{max}$ are periodic. 

Eq. \rf{1.5} was used in all computations. Convergence was checked on finer grids.
For small surface slopes the results provided by Eqs. \rf{1.5} and \rf{1.7}  are identical.

\begin{center}
{\bf III. RESULTS}
\end{center}

The cases of weak and strong anisotropy are addressed separately. 

\begin{center}
{\bf A. Zero or weak anisotropy}
\end{center}

Linearization of Eq. \rf{1.7} about the equilibrium solution $h=1$ gives
\begin{equation}
\xi_t = B\left(q\Lambda_1\xi_{xxxx}+\exp{\left(-1/r\right)}r^{-2}a_1\xi_{xx}\right),\quad \Lambda_1 < 0,
\label{1.9}
\end{equation}
where $\xi(x,t)=\exp{\left(\omega t + ikx\right)}$ is the perturbation of the equilibrium solution, $\omega$ is
the growth rate and $k$ is the wavenumber. Thus for the growth rate one obtains
\begin{equation}
\omega(k) = B\left(q\Lambda_1k^4-W_0k^2\right),
\label{1.10}
\end{equation}
where the notation is introduced $W_0=\exp{\left(-1/r\right)}r^{-2}a_1$. Obviously, $W_0 <0$ for $a_1<0$.
It follows from Eq. \rf{1.10} that 
the equilibrium surface is unstable ($\omega(k)>0$) to perturbations with the wavenumbers $0< k < k_c$ (where 
$k_c = (W_0/(Bq\Lambda_1))^{1/2}$), $\omega(0)=\omega(k_c)=0$ and,
in the interval $0\le k \le k_c$, $\omega(k)$ has the maximum value 
$\omega_{max} = -W_0^2/(4Bq\Lambda_1)$ at $k_{max} = k_c/\sqrt{2}$.
For $k\ge k_c$ the equilibrium surface is stable. Fig. 1 shows the sketch of $\omega(k)$.

Simulations are performed for $\epsilon_\gamma = 0$ and $1/18$, and $r=0.1,\ 1$. 
The graphs of the non-dimensionalized, reduced two-layer wetting chemical potential $\mu_w$ are shown in Fig. 2. 
For $r=0.1$, there is no substrate influence at the length scales larger than the unperturbed 
film thickness; for $r=1$, the substrate-film interaction matters at all length scales 
less than ten times the unperturbed film thickness.
For $r=1$, the reduced potential is less strong, but more uniform 
across the film.
Note that all figures in the paper 
show non-dimensional quantities.

When the anisotropy is zero, the film dewets for both strengths of the wetting potential, as expected
(other values of $r$, such as $r=0.01$ and $r=10$ also have been tried).
Thus for the remainder of the paper the focus is on the realistic
case of non-zero anisotropy, where a non-trivial evolution emerges. 

Fig. 3 shows the evolution of the surface shape for $\epsilon_\gamma = 1/18$ and $r=0.1$. 
This was obtained using Eq. \rf{1.5} (where $\Delta=0$),
with the initial amplitude $\alpha=0.01$ in Eq. \rf{ini_cond}. 
The shapes are smooth. 
Recalling the scaling, 
one can see that the 10 nm film completely dewets in approximately 45 seconds at 800 K.
This is consistent with the order of magnitude time scale in the experiments \cite{YZSLLHL,SECS}.
Fig. 4(a) shows $ln(1-h_{min})$ vs. time, where $h_{min}$ is the height of the surface at the tip 
of the pinhole. (Note that at $t=0$, $1-h_{min}=\alpha$). 
Clearly, the linear theory fails to correctly predict the growth rate after $t\approx 10^7$ (6 s.).
(Computing with the ``small-slope" Eq. \rf{1.7} instead of Eq. \rf{1.5} gives the same curve in Fig. 4(a),
but the maximum slope of the pinhole wall at the rupture time is three times larger in the case of Eq. \rf{1.7}.)

Fig. 5 shows the evolution of the surface shape for $r=1$. The transient shape approaches the equilibrium shape, 
which is different from the 
unstable trivial equilibrium $h=1$, and reaches this new equilibrium at $t\approx 2\times10^4$, i.e.
in a small fraction of a second.
Fig. 4(b) presents the growth rate data for this case. The equilibrium was confirmed by numerically
solving the steady-state (boundary value) fourth-order problem for Eq. \rf{1.7} (using the shooting method).
Most importantly, however,  for all values 
of $\alpha$ in $(0,0.99999]$ the initial deformation of the planar surface evolves to the shown equilibrium shape.
Thus the model predicts the absence of dewetting in this case. 


\begin{center}
{\bf B. Strong anisotropy}
\end{center}

Linearization of Eq. \rf{1.7} about the equilibrium solution $h=1$ gives
\begin{equation}
\xi_t = B\left(q\Lambda_1\xi_{xxxx}+\Delta\xi_{xxxxxx}+W_0\xi_{xx}\right),\quad 
\Lambda_1 \ge 0,\ W_0 < 0.
\label{1.11}
\end{equation}
The corresponding growth rate
\begin{equation}
\omega(k) = B\left(q\Lambda_1k^4-\Delta k^6-W_0k^2\right).
\label{1.12}
\end{equation}
The equilibrium surface is unstable ($\omega(k)>0$) to perturbations with the wavenumbers $0< k < k_c$, and 
$k_c^2 = q\Lambda_1/(2\Delta)+(q^2\Lambda_1^2/(4\Delta^2)-W_0/\Delta)^{1/2}$. For $k\ge k_c$ the equilibrium surface is stable.
The maximum growth rate and the
corresponding wavenumber can be easily obtained.

\begin{center}
\textit{1.} $\;\epsilon_\gamma = 0.0697$.
\end{center}

For this value of $\epsilon_\gamma$ (which is larger than the critical value 1/15 by only 4.5\%)
and $r=0.1$ or $r=1$ the evolution of the pinhole shape is qualitatively similar
to the previous case $\epsilon_\gamma = 1/18$ and $r=1$. For both values of $r$ the 
equilibrium and the transient 
shapes are smooth.

For $r=0.1$ the equilibrium shape has
the minimum at $z=0.92$. This equilibrium state attracts all perturbations with the initial amplitudes $\alpha$
in $(0,0.995)$. For $\alpha = 0.995$ or larger, the pinhole tip approaches the substrate until 
it touches it, i.e. the pinhole dewets. By comparing this case to Fig. 3, one can see that
the tiny supercritical anisotropy is almost successful in the suppression of dewetting.

For $r=1$, the equilibrium shape has the minimum at $z=0.95$,
and even for $\alpha=0.99999$ there is no dewetting.

\begin{center}
\textit{2.} $\;\epsilon_\gamma = 1/12$.
\end{center}

For $r=0.1$, there emerges a smooth equilibrium shape with the minimum at $z=0.97$, which attracts
the initial perturbations with the amplitudes $\alpha$ in $(0,0.5]$.
The transient shape exhibits smooth curved ``facets" and facet junctions. These facets gradually disappear
as the equilibrium shape is approached. 
If $\alpha$ is in $(0.5,0.99999]$, the transient shapes are truly faceted, and each shape evolves about the
initial shape, becoming more faceted with time. Thus again, for $\;\epsilon_\gamma = 1/12$ and $r=0.1$
there is no dewetting. 
Fig. 6 shows the initial and the final shapes for $\alpha=0.8$. 
Note that the tip of the pinhole is not faceted.
In the dimensional time, the shape changes in Fig. 6 
are practically instantaneous. After the low-energy orientations (0$^\circ$ and 90$^\circ$) have been formed the time steps required to track the evolution 
become very small,
and the computation is interrupted. (Of course, this is the drawback of the current model based on the evolution equation
for the surface height; better model would employ the parametric representation of the surface and will allow 
surface slopes of the 90$^\circ$ and larger \cite{KEdge}. I am presently working on such model.)

For $r=1$, the smooth equilibrium shape has the minimum at $z=0.965$.
This shape has a very narrow domain of attraction; already for $\alpha=0.1$, the clear tendency of the film to dewet
can be seen in 
Fig. 7.
The computation was interrupted after
the last shape in 
Figs. 7(a) and 7(b)
was dumped, since the time steps less than $10^{-10}$ are needed to compute further.
Such small time steps are warranted since the slope of the last shape near the tip is close to 90$^\circ$
and the evolution is very fast.
These Figures also suggest that as $\alpha$ increases, the tip of the pinhole elongates while the rest of the pinhole
shape changes little (is ``left behind"). In other words, the dewetting dynamics is confined to the tip of the pinhole.
Thus, unless the facets form on the transient shape (Fig. 7(a)),
a very narrow slit-like dewetting front is expected to extend to the substrate rapidly.
If the facets do form, then the tendency to dewet will be partially or completely suppressed.

\begin{center}
{\bf IV. DISCUSSION}
\end{center}

The complex scenarios of a thin solid film dewetting under the action of the two-layer
wetting potential have been computed using the nonlinear continuum model.
These scenarios depend on the strength of the surface energy anisotropy,
the characteristic wetting length and on the depth of the pre-existing pinhole defect in the film surface.

A film with the weak surface energy anisotropy dewets easily when the wetting potential 
is confined to the length scales of the order of the initial film thickness.
When the wetting potential extends beyond the initial film thickness, there is no dewetting but instead
the equilibrium emerges, where the pinhole is elongated but its tip does not touch the substrate.

When the anisotropy is strong, it is generally expected that the faceting prevails over dewetting.
For a very strong anisotropy either the smooth equilibrium emerges for shallow initial depths, 
or the transient shape becomes faceted
and the dewetting ceases.
Eqs. \rf{1.6c} and \rf{1.6d} make clear the reason for such behavior. Indeed, 
the dominant terms in $P^{(1)}_w$ ($P^{(2)}_w$)
are proportional to 
$h_x^{-1}$ ($h_x^{-7}$). Thus for large slopes the ``wetting" terms are small and they loose in the competition
with the anisotropic terms $P^{(1)}_\kappa$ and $P^{(2)}_\kappa$.


Two caveats must be noted. First,
the averaging employed for the calculation of $\mu_\kappa$ is the approximation
designed in order to obtain the simple model. 
Without averaging, the expression for the 
surface stiffness
contains the exponential terms. As a result, the expressions for
$P^{(1)}_\kappa$ and, especially, $P^{(2)}_\kappa$ in Eq. \rf{1.5} are cumbersome
and even the linear stability analysis for the strong anisotropy is complicated.
In Appendix the partial proof is presented 
that use of averaging does not change 
the results qualitatively and that
even the quantitative
impact on the dynamics due to averaging is very small.
Second, this model disregards the wetting stress \cite{LGDV}. However, 
as this reference suggests,
the impact of such stress is not strong (except near the rupture point, 
where the stress singularity is expected).  

\begin{center}
{\bf Acknowledgemets}
\end{center}

I thank Professor Brian J. Spencer for useful discussion.

\begin{center}
{\bf Appendix}
\end{center}

Here Eqs. \rf{1.4f}, \rf{1.2} and \rf{1.4b} are used as shown to derive the primary 
``wetting" part of 
$\mu_\kappa$ in the small slope approximation.
The contribution of wetting energy to the regularization term, $P^{(2)}_\kappa$ in Eq. 
\rf{1.8b}, is ignored. Thus in this extended SSA model Eqs. \rf{1.8b} - \rf{1.8d} are unchanged,
but $P^{(1)}_\kappa$ shown in Eq. \rf{1.8a} is augmented by the exponential term
\begin{eqnarray}
P^{(1)}_{\kappa,w} &=& \exp{\left(-h/r\right)}\left[-\left(\Lambda_1 + \Gamma\right)h_{xxx}+ \right. \nonumber \\
&& \left(r^{-1}\left(\Lambda_1 + \Gamma\right)-\Lambda_2h_{xx}+3\Gamma h_{xx}\right)h_{xx}h_x + \nonumber \\
&& \left.\left(2\Gamma-\Lambda_3\right)h_{xxx}h_x^2 + r^{-1}\left(\Lambda_3-2\Gamma\right)h_{xx}h_x^3\right].
\label{appn}
\end{eqnarray}
Also, since averaging is not employed, $q$ is set equal to one in Eq. \rf{1.7}.

For the basic and the exdended models
Fig. 8 
compares the dewetting rates in the case $\epsilon_\gamma=1/18$ and $r=0.1$.
This case was chosen because it is dominated by
``wetting" contributions. The former rate is also shown in Fig. 4(a). Clearly the results are identical. For the large slope cases such as shown in
Figs. 6 and 7, the differences are expected very small, if any, since the wetting correction to
$P^{(1)}_{\kappa}$ in Eq. \rf{1.6a} is proportional to $h_x^{-3}$ (vs. the $h_x^{-1}$ dependence in Eq. \rf{1.6c}).

\newpage

\begin{center}
{\bf FIGURES CAPTIONS}
\end{center}

\noindent
FIG. 1. Sketch of the linear growth rate $\omega(k)$. Perturbations with wavenumbers
$0< k < k_c$ are unstable and may grow nonlinearly until the film ruptures.
\vspace*{0.5cm}\\
\noindent
FIG. 2. The reduced nondimensional wetting chemical potential $\mu_w=-(1/r)(\Gamma - 1)\exp{(-h/r)}$.
This formula is obtained when $\gamma_f(\theta)$ is taken isotropic and $h_x$ is taken zero
in Eq. \rf{1.4e}. $\Gamma = 0.5$.
Solid line: $r=0.1$; Dashed line: $r=1$.
\vspace*{0.5cm}\\
\noindent
FIG. 3. Surface shapes for $r=0.1$ and weak anisotropy ($\epsilon_\gamma=1/18$). 
\vspace*{0.5cm}\\
\noindent
FIG. 4. (a) Growth rate data for Fig. 3. Slope of the solid line equals the growth rate of the instability. 
For comparison, the dashed line is $\ln{\alpha} + \omega_{max}t$, where $\omega_{max}$ is the linear growth rate.
(b) Same as (a), but for Fig. 5.
\vspace*{0.5cm}\\
\noindent
FIG. 5. Surface shapes for $r=1$ and weak anisotropy ($\epsilon_\gamma=1/18$). 
\vspace*{0.5cm}\\
\noindent
FIG. 6. Transient faceting of the surface for strong anisotropy. $\epsilon_\gamma = 1/12,\ r=0.1$ and $\alpha=0.8$. 
\vspace*{0.5cm}\\
\noindent
FIG. 7. Transient surface shapes for $\epsilon_\gamma = 1/12,\ r=1$.
(a) $\alpha=0.1$. (b) $\alpha=0.25$.
\vspace*{0.5cm}\\
\noindent
FIG. 8. Comparison of the dewetting rates for $\epsilon_\gamma=1/18$ and $r=0.1$.
Solid line: basic SSA model with averaging, given by Eqs. \rf{1.7} - \rf{1.8d}. Dashed line: extended 
SSA model as described in Appendix.

\newpage

\begin{figure}[!t]
\includegraphics[width=6.5in]{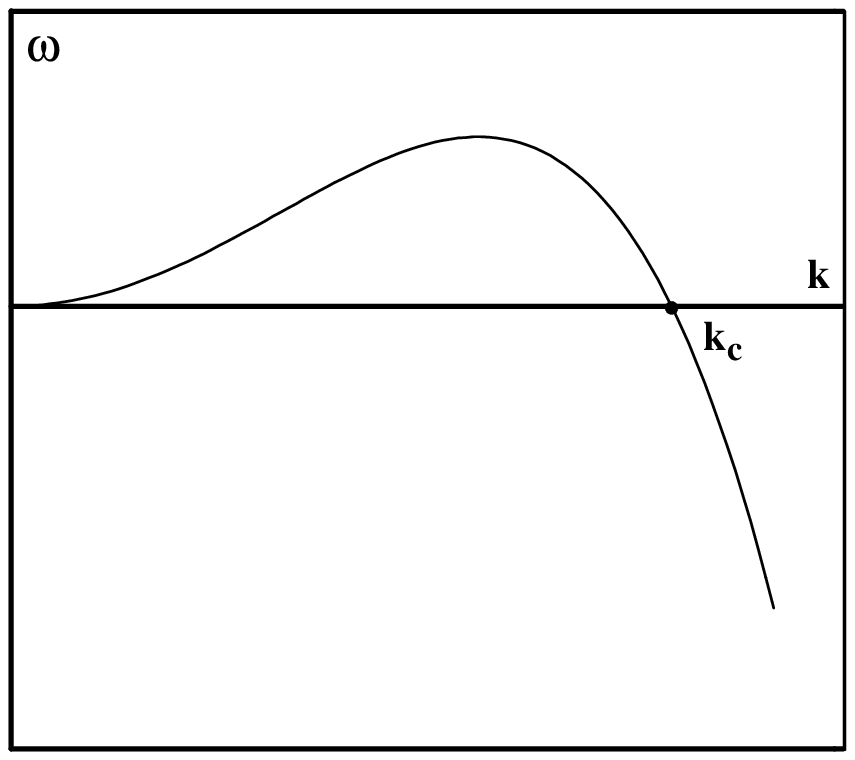}
\caption{} \label{fig:1}
\end{figure}

\newpage

\begin{figure}[!t]
\includegraphics[width=6.5in]{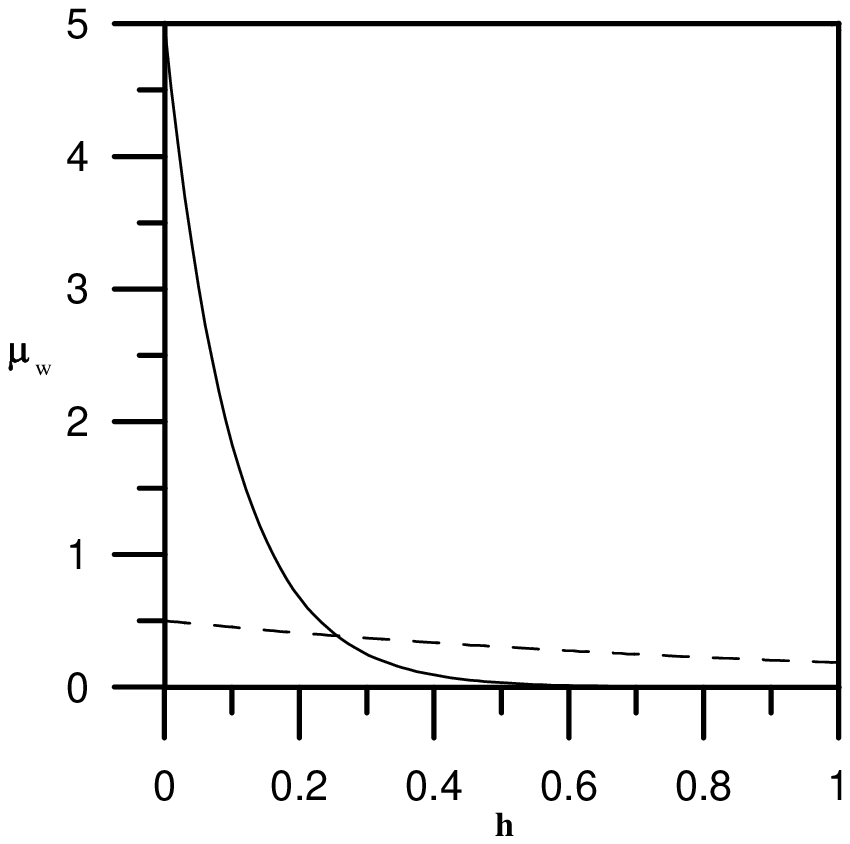}
\caption{} \label{fig:Vz}
\end{figure}

\newpage

\begin{figure}[!t]
\includegraphics[width=6.5in]{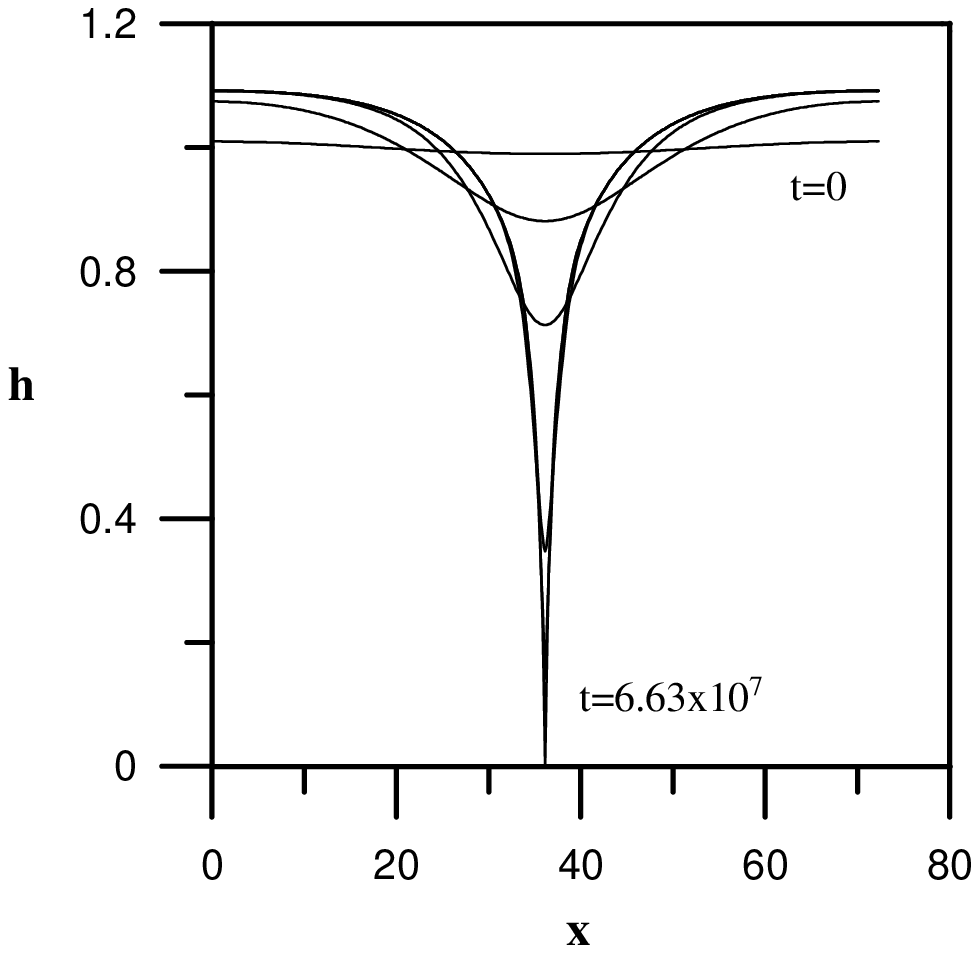}
\caption{} \label{fig:Vm}
\end{figure}

\newpage

\begin{figure}[!t]
\includegraphics[width=6.5in]{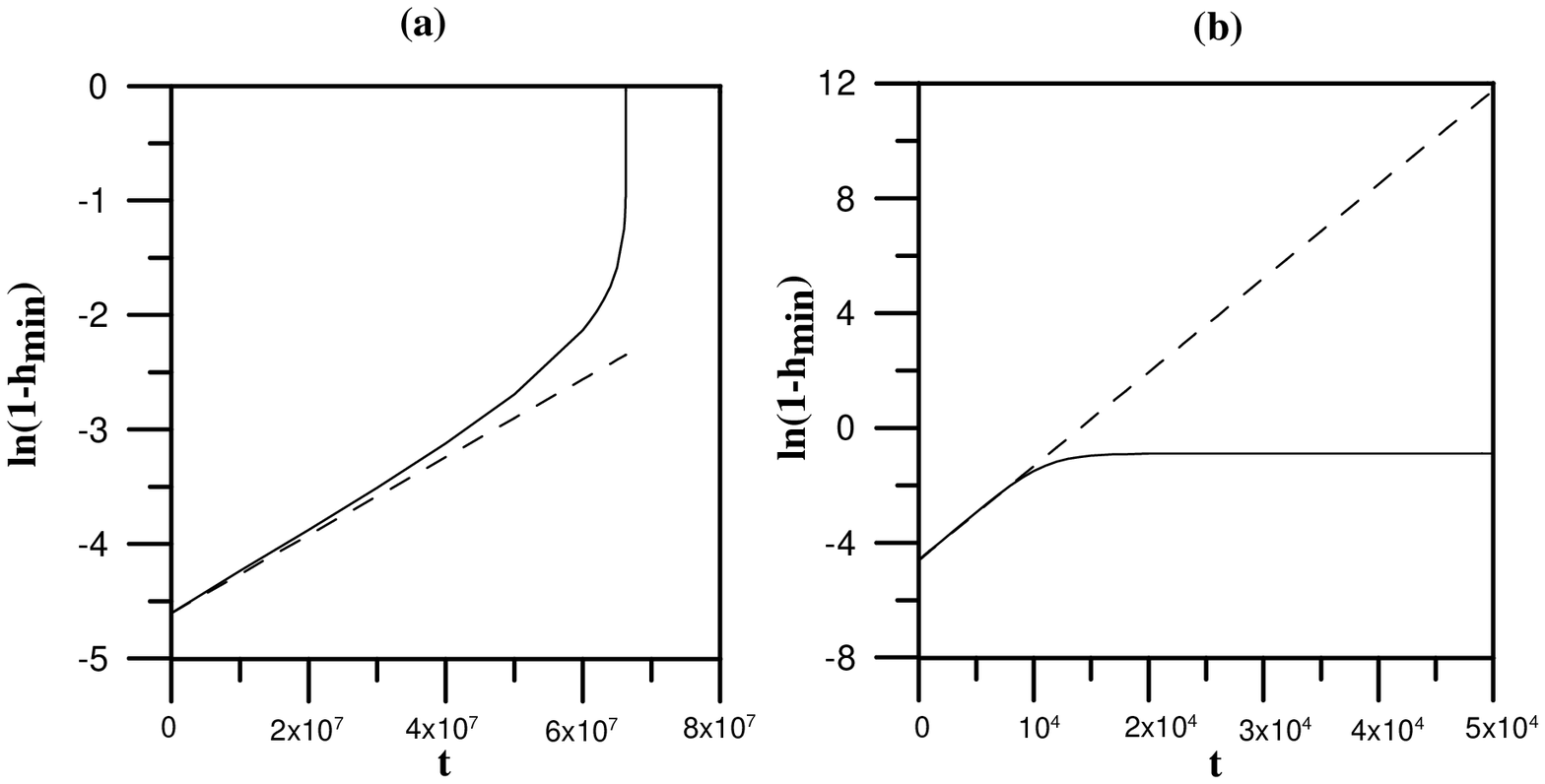}
\caption{} \label{fig:Om1}
\end{figure}

\newpage

\begin{figure}[!t]
\includegraphics[width=6.5in]{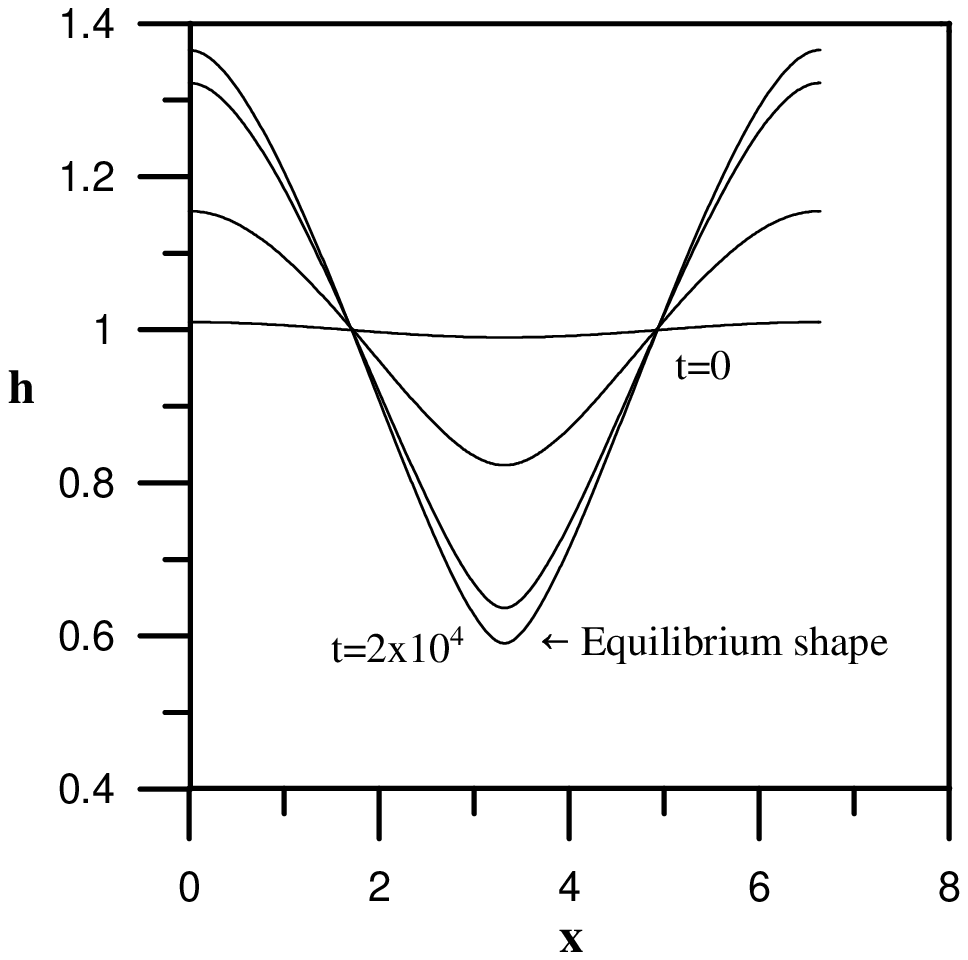}
\caption{} \label{fig:Om10}
\end{figure}

\newpage

\begin{figure}[!t]
\includegraphics[width=6.5in]{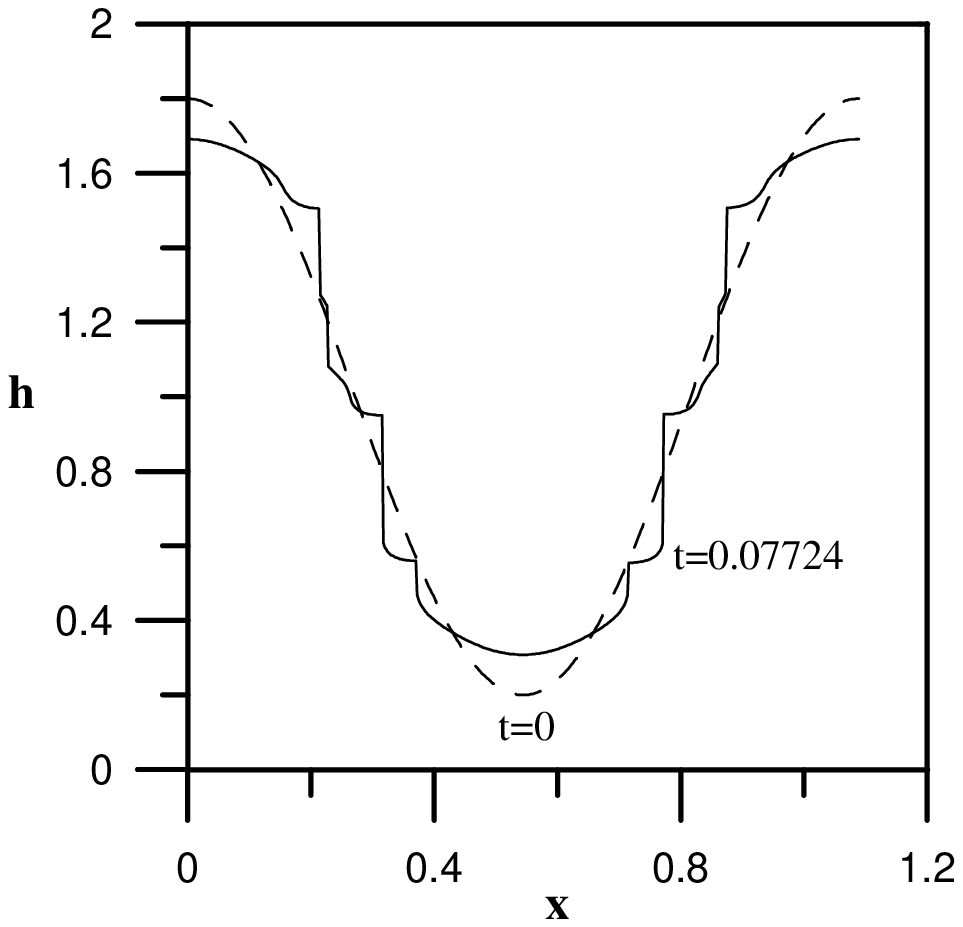}
\caption{} \label{fig:Qf}
\end{figure}

\newpage

\begin{figure}[!t]
\includegraphics[width=6.5in]{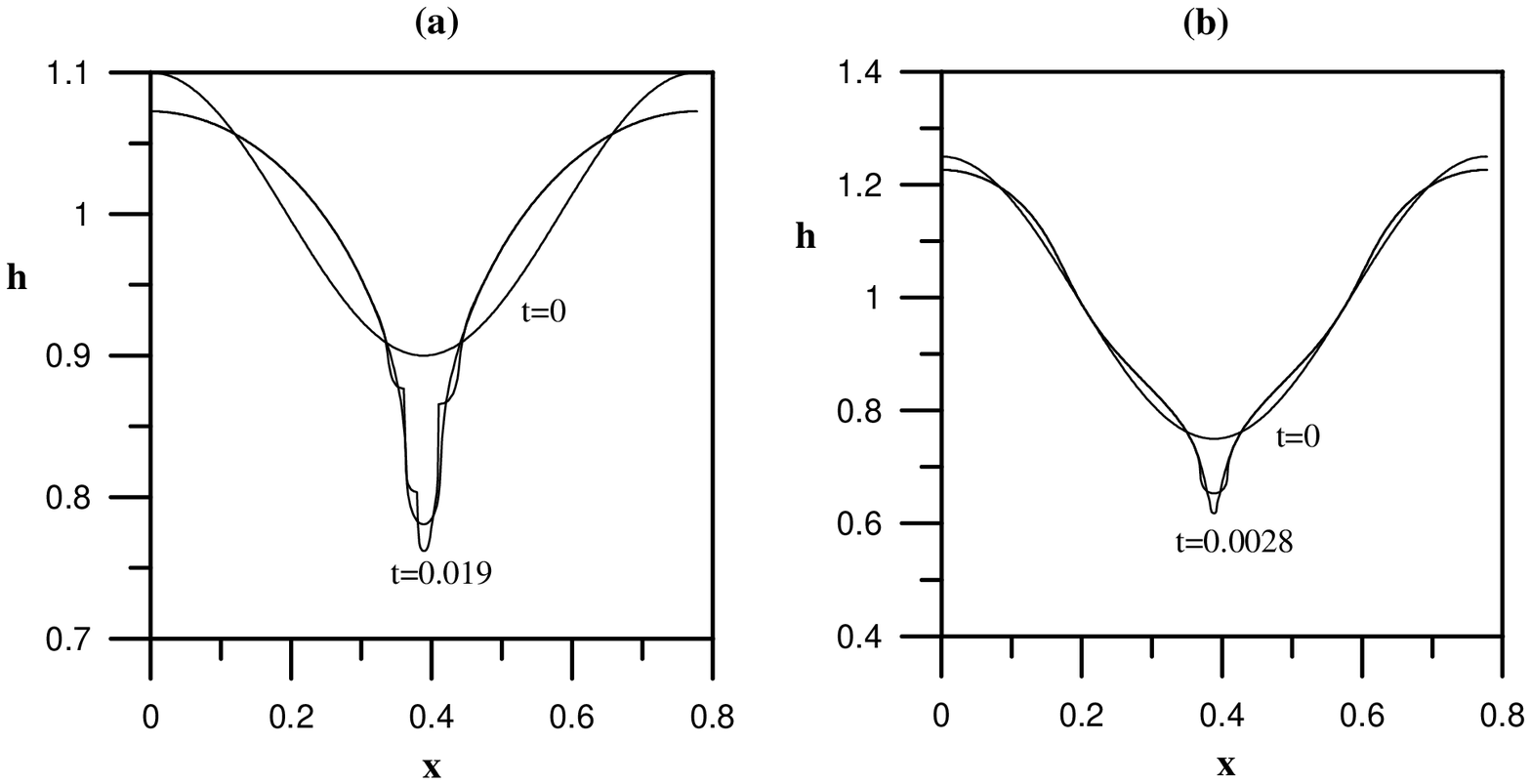}
\caption{} \label{fig:decay}
\end{figure}

\newpage

\begin{figure}[!t]
\includegraphics[width=6.5in]{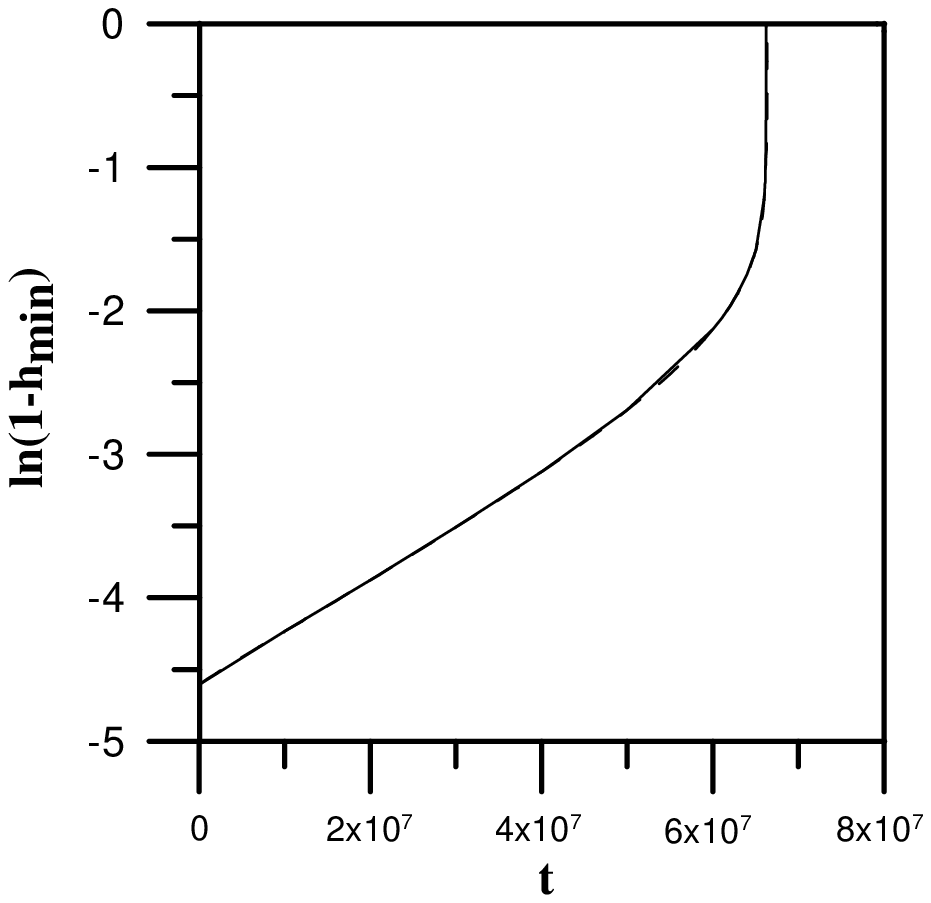}
\caption{} \label{fig:kc}
\end{figure}

\end{document}